\documentclass[aps,pre,letterpaper,twocolumn,showpacs]{revtex4-1}
\usepackage[utf8]{inputenc}
\usepackage{multirow}
\usepackage{amsmath,amsfonts,amssymb}
\usepackage{graphicx}
\usepackage{xcolor}

\def\be{\begin{equation}}
\def\ee{\end{equation}}
\def\ba{\begin{eqnarray}}
\def\ea{\end{eqnarray}}

\def\correction#1{#1}       

\begin{document}
\title{Tensor Renormalization-Group study of the surface
critical behavior of a frustrated two-layer Ising model}
\author{Christophe Chatelain}
\affiliation{Universit\'e de Lorraine, CNRS, LPCT, F-54000 Nancy, France}
\date{\today}

\begin{abstract}
Two replicas of a 2D Ising model are coupled by frustrated spin-spin
interactions. It is known that this inter-layer coupling is marginal
and that the bulk critical behavior belongs to the Ashkin-Teller (AT)
universality class, as the $J_1-J_2$ Ising model.
In this work, the surface critical behavior is
studied numerically by Tensor Renormalization-Group calculations.
The Bond-Weight Tensor Renormalization Group algorithm is extended
to tackle systems with boundaries. It is observed that the two-fold
degeneracy of the surface magnetic scaling dimension of the AT model
is lifted in the frustrated two-layer Ising model (F2LIM). The
splitting is explained by the breaking of the ${\mathbb Z}_2$-symmetry
under spin reversal of a single Ising replica in the F2LIM.
The two distinct surface magnetic scaling dimensions $x_1^s$ and $x_2^s$
of the F2LIM satisfies a simple duality relation $x_1^s=1/4x_2^s$.
\end{abstract}
\maketitle

\section{Introduction}
The frustrated $J_1-J_2$ Ising model has recently attracted a renewed
interest. Several Tensor Renormalization-Group calculations~\cite{Li1,
Yoshiyama,Gangat} questioned well-established results obtained earlier
by means of Monte Carlo simulations~\cite{Kalz1,Jin1,Kalz2,Jin2}.
In this context, we constructed a frustrated two-layer Ising model (F2LIM)
model sharing the same scaling limit as the $J_1-J_2$ Ising model and therefore
the same bulk critical behavior~\cite{Chatelain}. This model has the advantage
of being well suited to Tensor Renormalization-Group calculations. Using the
Bond-Weight Tensor-Renormalization Group (BTRG) algorithm~\cite{BTRG}, we
have shown that the critical behavior of this model belongs to the universality
class of the Ashkin-Teller (AT), as claimed by Monte Carlo simulations for the
$J_1-J_2$ Ising model. Like the 4-spin interaction of the AT model, the
inter-layer coupling of the F2LIM is marginal. Critical exponents therefore
vary along the critical line~\cite{Kadanoff,Nienhuis,Baxter} and the central
charge of the model is $c=1$.
\\

In this work, we extend the study of the F2LIM model to its surface critical
behavior. The latter is expected to be richer than the bulk one~\cite{Binder,
Cardy1,Pleimling}. It is indeed well known that several surface universality
classes may exist for a given bulk universality class. Even for the different
transitions (ordinary, special, extraordinary and surface transitions)
undergone by the same model at its surfaces, different universality classes
may exist. It is notably the case for the ordinary and special surface
transitions of the 3D Ising model~\cite{Pleimling}. From the point of
view of Conformal Field Theory (CFT), the surface critical behavior is
described by unitary irreducible representations of a single Virasoro
algebra rather than two, holomorphic and anti-holomorphic, in the
bulk~\cite{Cardy84,Cardy86a,Cardy86b,Henkel}. With Fixed Boundary
Conditions, only primary operators and their descendents compatible
with the boundaries appear in the spectrum. The critical behavior of the
AT model with Open Boundary Conditions (OBC) has been studied by numerical
diagonalization of the transfer matrix of small strips~\cite{Gehlen86,Gehlen87}
and by Bethe-ansatz calculations for an equivalent XXZ model~\cite{Alcaraz87}.
An interesting question is therefore whether the same spectrum is
observed for the F2LIM. Tensor Renormalization-Group algorithms
may help to address this question by allowing for larger system
sizes~\cite{Ueda,Ueda2,Huang,Guo}.
\\

The plan of this paper is the following. In Section II, our extension
of the BTRG algorithm to systems with boundaries is presented. The
estimation of scaling dimensions with the gap-exponent relation is
discussed at the end of this section. The case of the AT model is
first considered in Section III. Estimates of the lowest scaling
dimensions with OBC and identical and mixed FBC are presented.
Section IV is devoted to the study of the surface critical behavior
of the F2LIM. A new estimate of the critical line is obtained from
the behavior of the latent heat of a domain wall induced by mixed
Boundary Conditions. Scaling dimensions are estimated along this line
and compared with those of the AT model. Conclusions follow.

\section{Numerical details}
\subsection{Boundary BTRG algorithm}
\label{BTRG}
\begin{figure}
    \centering
    \includegraphics[width=0.49\textwidth]{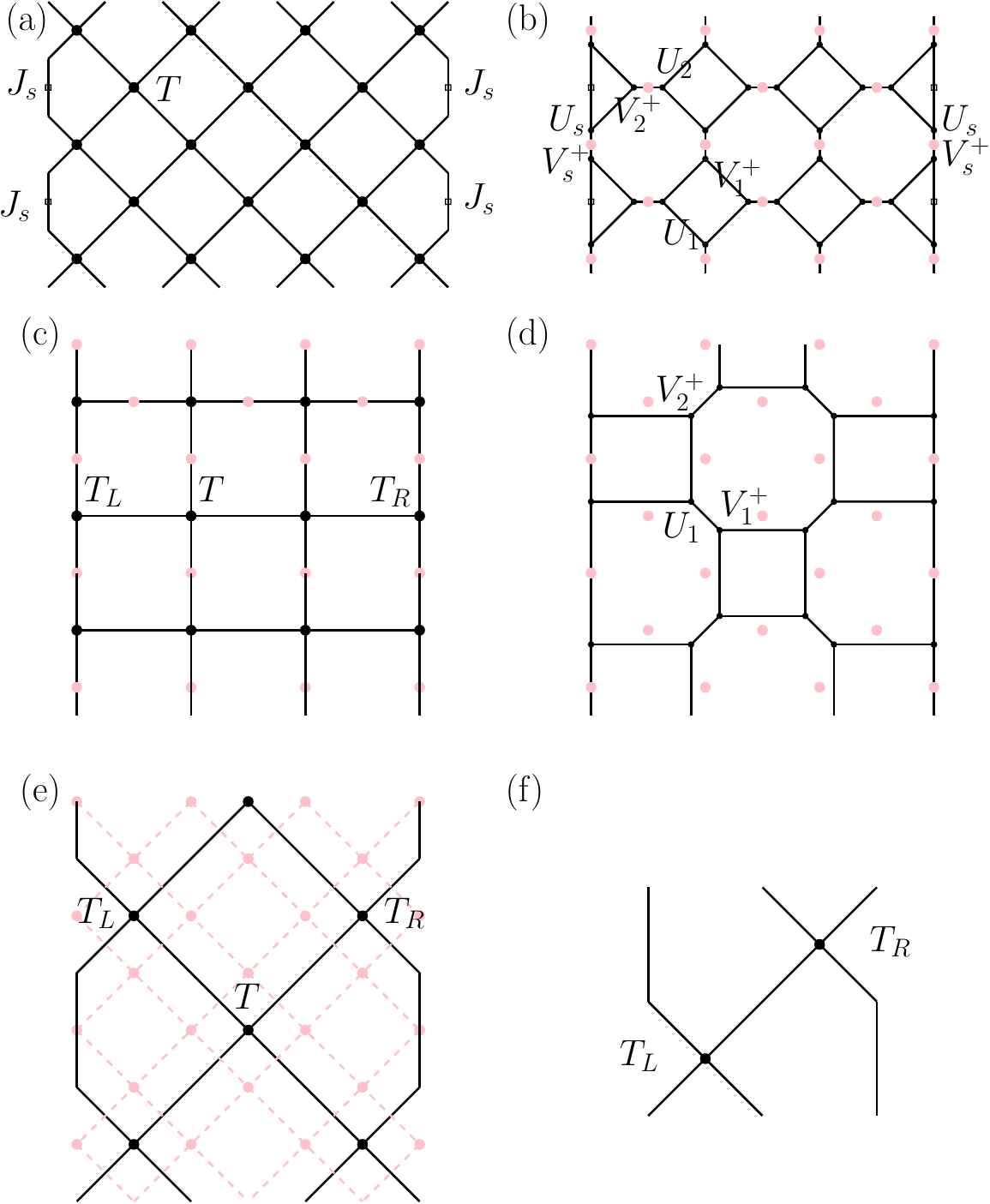}
    \caption{The different steps of the Tensor-Renormalization Group algorithm
    for a finite strip. The black dots correspond to the tensors. To each line
    is attached an index which is one of the indices of the tensors at the two
    edges of the line. A black dot at the crossing of four lines is therefore
    a rank-4 tensor. A summation over the index of each line is implicit.
    The pink dots correspond to the initial tensors. After the two steps
    of the TRG algorithm, one can see that the number of tensors has
    decreased by a factor of 4.}
    \label{fig0}
\end{figure}

The calculations were performed by means of a Tensor Renormalization-Group
(TRG) algorithm~\cite{Levin,Xiang} adapted to a system with boundaries.
Such a modification was recently presented for the Higher-Order Tensor
Renormalization-Group (HoTRG) algorithm~\cite{Ilino} and the Tensor-Network
Renormalization (TNR)~\cite{Ilino2}. Our algorithm is depicted on
Fig.~\ref{fig0} for a general 2D classical spin model on a square lattice.
The Boltzmann weight of a given spin configuration $\{s_i\}_i$ is first
expressed as a product of rank-4 tensors:
    \be e^{-\beta{\cal H}[s]}=T_{s_is_js_ks_l}T_{s_ns_ms_ps_q}\ldots\ee
In our setup, each tensor corresponds to the Boltzmann weight of a square
plaquette. The four indices of the tensor are the four spin variables at
the corners of the plaquette. In the presence of two boundaries, at the left
and the right of the system, it is necessary to add rank-2 tensors associated
to the Boltzmann weight of a coupling between two spins at the surface.
These tensors are denoted as $J_s$ on the subplot (a). During the first step
of the algorithm, tensors $T$ are recasted as a matrix and then decomposed
into two rank-3 tensors by a Singular Value Decomposition (SVD):
    \be T_{s_1s_2;s_3s_4}=\sum_n U_{s_1s_2;n}\sqrt{\Lambda_n}\times
    \sqrt{\Lambda_n}(V^+)_{n;s_3s_4}      \label{SVD}\ee
Depending on how the tensor was recasted, either as $T_{s_1s_2;s_3s_4}$
or $T_{s_4s_1;s_2s_3}$, the new rank-3 tensors are aligned
vertically or horizontally as can be seen on the subplot (b). To avoid
an exponential increase of the size of the tensors, the sum in Eq.~(\ref{SVD})
is truncated to keep only the $\chi$ largest singular values
$\Lambda_n$. This choice minimizes the error if defined as the
Hilbert-Schmidt norm
    \be\Big|\Big|T-\sum_{n=1}^\chi U_{.;n}\Lambda_n V_{n;.}^+\Big|\Big|_2,\quad
    ||A||_2=\sum_{ij} A_{ij}^2\ee
On a boundary, the tensor can only be decomposed into two vertically-aligned
rank-3 tensors. Therefore, one needs to distinguish between 4 pairs of
rank-3 tensors, two in the bulk and one at left and right boundaries respectively.
In the bulk, the lattice is now formed of plaquettes made of either 4 or 8
tensors. Contracting the plaquettes formed of 4 tensors yields new bulk
rank-4 tensors $T$. At the two boundaries, subplot (b) shows the presence of
plaquettes formed of 3 tensors.
By contracting them, one is left with new rank-3 tensors $T_L$ on the left
boundary and $T_R$ at the right one. These new tensors form a square lattice
at $45^\circ$ of the original one (subplot (c)). In the second step of the
algorithm, the bulk tensor $T$ is again decomposed in two rank-3 tensors
by a SVD. Plaquettes of 4 and 8 tensors are formed. The contraction of the
square plaquettes yields a new bulk rank-4 tensor $T$. At the two boundaries,
no decomposition of the tensors $T_L$ and $T_R$ is needed since there are
already square plaquettes that can be immediately contracted to give new
left and right rank-4 boundary tensors. Note that the boundary tensors
are connected by lines, meaning that the rank-2 tensor $J_s$ has been
replaced by an identity matrix after the first iteration of the algorithm.
The number of tensors is divided by two at each iteration of the
algorithm. Equivalently, after $n$ iterations, each tensor corresponds
to the Boltzmann weight of a block of $2^{n+1}\times 2^{n+1}$ spins.

\begin{figure}
    \centering
    \includegraphics[width=0.49\textwidth]{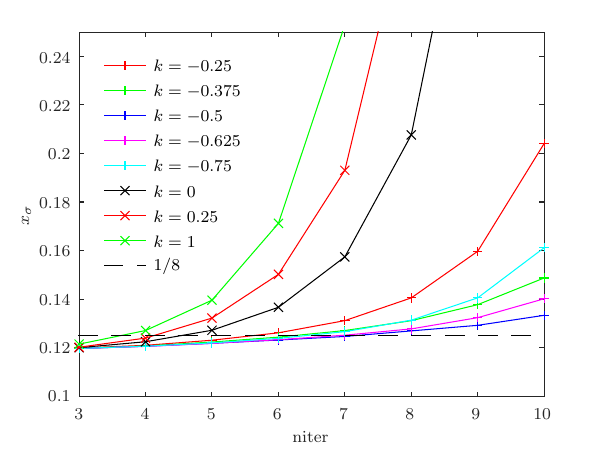}
    \caption{\correction{
    Lowest bulk scaling dimension of the F2LIM versus the number of
    iterations of the BTRG algorithm for different values of the parameter $k$.
    The calculation has been performed on the critical line of the model
    for $J_1=0.1344575$ and with $\chi=64$ states. The dashed line corresponds
    to the expected value $1/8$.}}
    \label{fig16}
\end{figure}

The accuracy of the estimates of the scaling dimensions is significantly
improved without any increase of the computational time with a variant of
the TRG algorithm, called the Bond-Weight Tensor-Renormalization Group
(BTRG)~\cite{BTRG}. Instead of decomposing the rank-4 tensors according
to Eq.~(\ref{SVD}), the SVD is written as
    \be T_{s_1s_2;s_3s_4}=\sum_n U_{s_1s_2;n}(\Lambda_n)^{1-k/2}\ \!
    (\Lambda_n)^k\ \!(\Lambda_n)^{1-k/2}(V^+)_{n;s_3s_4}      \label{SVD2}\ee
where $k$ is a parameter. The $k$-th power of the singular values
$(\Lambda_n)^k$ can be casted as a diagonal $\chi\times\chi$ matrix
that is inserted on each line joining two rank-3 tensors while the terms
$(\Lambda_n)^{1-k/2}$ are absorbed into the rank-3 tensors $U$ and $V^+$.
The TRG algorithm corresponds to the choice $k=0$ and it has been argued
for the Ising model that the optimal value is $k=-1/2$~\cite{BTRG}.
\correction{
As can be seen on Fig~\ref{fig16} for the F2LIM, the most stable estimate
of the lowest bulk scaling dimension is obtained for $k=-1/2$. One can
notice a large deviation from the expected value only after 8-10 iterations
of the BTRG algorithm. In the following, the parameter $k=-1/2$ will be
used and 10 iterations of the BTRG algorithm will be performed.}

\begin{figure}
    \centering
    \includegraphics[width=0.49\textwidth]{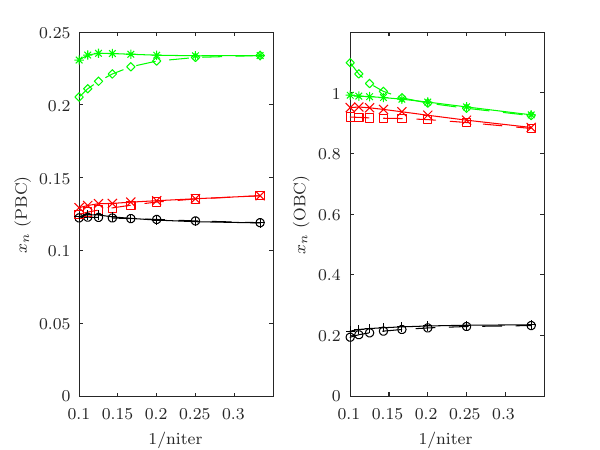}
    \caption{\correction{
    Bulk (left) and surface (right) scaling dimensions of the F2LIM at the
    critical point for which $J_0=J_1+J_2=0.3042435$ versus the number of
    iterations of the BTRG algorithm. The dashed lines and open symbols
    were computed with $\chi=32$ states and the solid lines and crosses
    with $\chi=64$ states.}}
     \label{fig17}
\end{figure}

\correction{
Tensor-network Renormalization-Group algorithms rely on a variational approach.
The accuracy of the method is expected to be improved by increasing the number
of states $\chi$. Nevertheless, it is difficult to estimate the deviation of the
scaling dimensions from the exact values when the latter are not known.
On Fig.~\ref{fig17}, the bulk and
surface scaling dimensions of the F2LIM at the point $J_0\simeq 0.3042$ of the
critical line are plotted for $\chi=32$ and $64$ states. For the lowest
scaling dimension, the difference between the two is small, especially in
the bulk. However, an important deviation is observed for the third scaling
dimension. The latter is more stable for $\chi=64$ while, for $\chi=32$,
it tends to either decrease or increase rapidly as the number of iterations
of the BTRG algorithm grows. In the following, the scaling dimensions will
be computed with $\chi=64$ states.}

\subsection{Estimate of the scaling dimensions}
\label{sec:scal_dim}
The lowest bulk and surface scaling dimensions of the spin model can be
estimated by constructing two transfer matrices, one with Periodic Boundary
Conditions (PBC) using the bulk tensor $T$, and the second with Open (OBC)
or Fixed Boundary Conditions (FBC) by a product of the boundary tensors
$T_L$ and $T_R$. In the second case, the transfer matrix is shown on the subplot
(f) of Fig.~\ref{fig0}. The largest eigenvalues $\Lambda_0>\Lambda_1>\ldots
>\Lambda_n$ of these transfer matrices are computed with the Arnoldi algorithm
of the {\tt arpack} library~\cite{arpack}. The scaling dimensions are extracted
from the energy gaps $E_n-E_0=\ln{\Lambda_n/\Lambda_0}$ using the gap-exponent
relation~\cite{Cardy84b}
    \be x_n={L\over 2\pi}\big(E_n-E_0\big)  \label{GapExponentPBC}\ee
in the limit of a number of spins $L$ going to infinity. Irrelevant scaling
operators introduce algebraic corrections to these relations. Marginal
operators, as in the 4-state Potts model, yields logarithmic
corrections~\cite{Cardy86}
    \be x_n={L\over 2\pi}\big(E_n-E_0\big)+{d_n\over\ln L}.
    \label{GapExponentLog}\ee
In the framework of conformal invariance, the scaling dimensions
are associated to primary operators ($x_n=\Delta+\bar\Delta$) and
their descendants ($x_n=\Delta+\bar\Delta+n+\bar n$ with $n,\bar n\in
\mathbb{N}^*$). With Open or Fixed Boundary Conditions, the gap-exponent
relation reads
    \be x_n^s={L\over\pi}\big(E_n-E_0\big)  \label{GapExponentFBC}\ee
because the strip is mapped onto the upper half of the complex plane only.

\section{Surface critical behavior of the Ashkin-Teller model}
\subsection{The Ashkin-Teller model}\label{SecIIA}
The classical Ashkin-Teller model corresponds to two Ising replicas,
denoted $\sigma^A$ and $\sigma^B$ in the following, coupled by their
local energy. The Hamiltonian reads
    \ba -\beta H&=&J_2
    \sum_{i,j}\sigma^A_{i,j}\big[
    \sigma^A_{i+1,j}+\sigma^A_{i,j+1}\big]\nonumber\\
    &+&J_2\sum_{i,j}\sigma^B_{i,j}\big[
    \sigma^B_{i+1,j}+\sigma^B_{i,j+1}\big]\nonumber\\
    &+&J_1\sum_{i,j}\sigma^A_{i,j}\sigma^B_{i,j}\big[
    \sigma^A_{i+1,j}\sigma^B_{i+1,j}
    +\sigma^A_{i,j+1}\sigma^B_{i,j+1}\big]
    \label{Eq1}
    \ea
with $\sigma^A_{i,j},\sigma^B_{i,j}=\pm 1$. This Hamiltonian is invariant
under a global spin-flip of each replica, i.e. $\sigma^{A_{i,j}}
\longrightarrow -\sigma^{A_{i,j}}$ and $\sigma^{B_{i,j}}\longrightarrow
-\sigma^{B_{i,j}}$, and under the exchange $\sigma^{A_{i,j}}
\longleftrightarrow \sigma^{B_{i,j}}$. The symmetry group is therefore
the dihedral group $\mathbb{D}_4$. The model reduces to two uncoupled
Ising models when $J_1=0$ (symmetry group $\mathbb{Z}_2\otimes \mathbb{Z}_2$)
and to the 4-state Potts model when $J_1=J_2$ (symmetry group $\mathbb{S}_4$).
Between these two lines of the phase diagram, the Ashkin-Teller model is
critical along the self-dual line
    \be e^{-2J_1}=\sinh 2J_2\ee
Remarkably, the coupling $J_1$ is marginal, inducing a continuous variation
of the critical exponents along the critical line. Exploiting a mapping onto
the exactly-solved 8-vertex model and the Coulomb gas, the critical
exponents have been determined exactly~\cite{Kadanoff,Nienhuis,Baxter,Yamanaka}
    \be x_{\sigma}={1\over 8},\quad
    x_{\sigma\tau}={1\over 8-4y},\quad
    x_\varepsilon={1\over 2-y}=4x_{\sigma\tau},\quad
    \label{ScalDimAT}\ee
where the parameter $y$ is in the range $[0;3/2]$ along the critical line
and is related to the coupling $J_1$ by
    \be\cos{\pi y\over 2}={1\over 2}\Big[e^{4J_1}-1\big].\ee
The critical exponents of the Ising and the 4-state Potts models are indeed
recovered for $y=1$ and $y=0$ respectively. The model is also conformally
invariant with a central charge $c=1$.
\\

\correction{
As mentioned above, the Ashkin-Teller model is equivalent to the 4-state Potts
model when $J_1=J_2$, The $q$-state Potts model undergoes a second-order phase
transition for $q\le 4$ and a first-order one for $q>4$. In early
real-space RG studies of this model, the fugacity of vacancies was
shown to be marginal at $q=4$ and relevant for $q>4$~\cite{Nienhuis79}.
The phenomenological RG equations proposed by Cardy {\sl et al.}~\cite{Cardy80}
yield logarithmic corrections that were later observed in Monte Carlo
simulations~\cite{Salas}. Such logarithmic corrections are also expected
at the tricritical point of the Ashkin-Teller model. The precise form of
the logarithmic corrections in the 4-state Potts model is still an
open question~\cite{Berche}.}
\\

The surface critical exponents of the Ashkin-Teller model with Open
Boundary Conditions have been determined by exact diagonalization
of transfer matrices on small strips~\cite{Gehlen86,Gehlen87}. Exact values
have been conjectured based on a numerical solution of the Bethe equations
of a quantum XXZ chain with complex boundary fields~\cite{Alcaraz87}:
    \be x^s_\sigma=(2x_\varepsilon)^{-1},\quad
    x^s_{\sigma\tau}=1,\quad
    x_\varepsilon^s=2.
    \label{ScalSurfDimAT}\ee
For the Ising model, all the eigenvalues of the transfer matrix have been
determined exactly for various BCs, including OBC and FBC~\cite{OBrien}.
From the first gap, one can show that the smallest scaling dimension
is $x_1^s=2$ for $(++)$ BCs and 1 for $(+-)$ BCs~\cite{Izmailian}.

\subsection{Scaling dimensions of the AT model}

\begin{figure}
    \centering
    \includegraphics[width=0.49\textwidth]{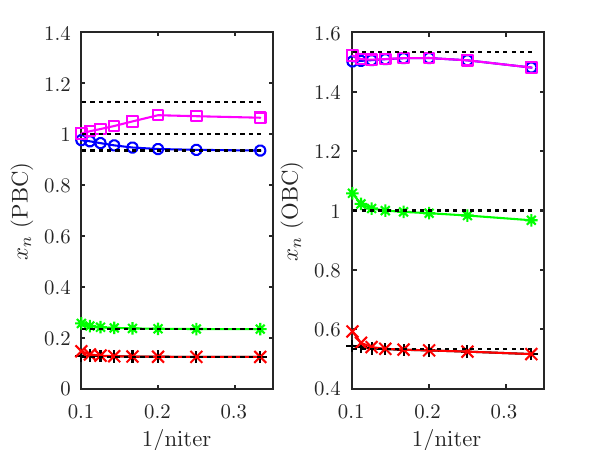}
    \caption{Estimates of the bulk (left) and surface (right) critical
    dimensions of the Ashkin-Teller model versus the inverse of the number of
    iterations of the BTRG algorithm at the point $J_1=0.04894351$ (close to
    the Ising point) of the critical line. The dashed lines are the exact
    values.}
    \label{fig1}
\end{figure}

\begin{figure}
    \centering
    \includegraphics[width=0.49\textwidth]{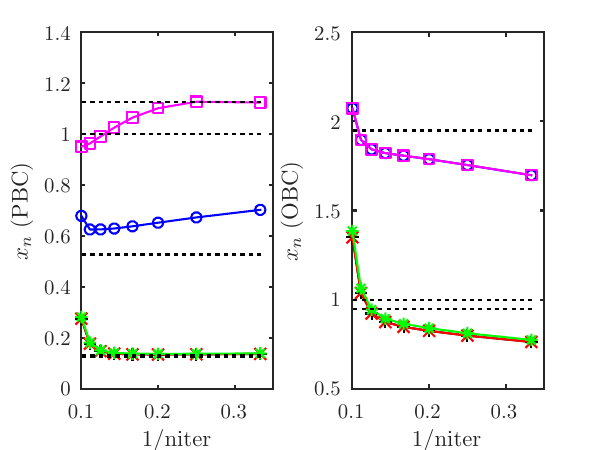}
    \caption{Estimates of the bulk (left) and surface (right) critical
    dimensions of the Ashkin-Teller model versus the inverse of the number of
    iterations of the BTRG algorithm at the point $J_1=0.2724481$ (close to the
    4-state Potts point) of the critical line. The dashed lines are the
    exact values.}
    \label{fig2}
\end{figure}

As explained in~\ref{sec:scal_dim}, the scaling dimensions are estimated
using the gap-exponent relation Eq.~(\ref{GapExponentPBC}). However, the latter
holds only in the limit of an infinite strip width $L$. As a consequence,
the estimates obtained after a finite number of BTRG iterations are
expected to deviate from the expected asymptotic values due to Finite-Size
effects. Moreover, systematic deviations are also expected at large number
of iterations because of the accumulation of errors at each Renormalization
step. It is therefore necessary to find an intermediate number of iterations
where the effective scaling dimensions are stable. As can be seen on Fig.~\ref{fig1}
for a small value of $J_1$, i.e. close to the Ising point, the estimates of
the three smallest bulk scaling dimensions shows a well-defined plateau.
A small systematic shift of the fourth scaling dimension from the exact value
is observed at large numbers of iteration while the fifth scaling dimension
deviates significantly from the exact value. Measuring the scaling dimensions
after 6 or 7 iterations appears to be a reasonable choice at least for the
four smallest scaling dimensions. This choice leads to stable surface
scaling dimensions too. However, at larger values of $J_1$, i.e. close to the
4-state Potts point, the scaling dimensions display a strong dependence
on the number of iterations (Fig.~\ref{fig2}).
\correction{As discussed in~\ref{SecIIA}, logarithmic corrections are
expected at the 4-state Potts point of the AT model. They manifest
themselves as a logarithmic correction in the gap-exponent relation
as given by Eq.~(\ref{GapExponentLog}).} Since the strip width $L$ grows
exponentially with the number $n_{\rm iter}$ of BTRG iterations, the
correction is expected to scale as $1/(n_0+n_{\rm iter})$. In practise,
fitting this correction is a complicated task because level crossings
often occur. In the following, we will restrict ourselves to consider
the estimates of the scaling dimensions after 7 BTRG iterations. Additional
systematic deviations are therefore expected close to the 4-state Potts point.

\begin{figure}
    \centering
    \includegraphics[width=0.49\textwidth]{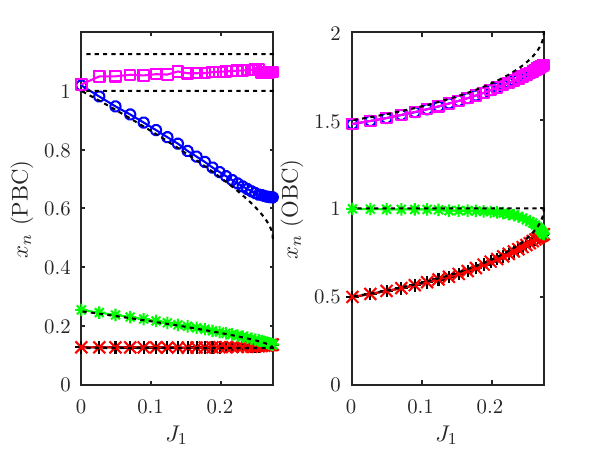}
    \caption{Estimates of the bulk (left) and surface (right) critical
    dimensions of the Ashkin-Teller model versus the coupling $J_1$.
    The dashed lines are the exact values.}
    \label{fig3}
\end{figure}

As can seen on the left subplot of Fig~\ref{fig3}, the bulk critical exponents
are in good agreement with the expected values Eq.~(\ref{ScalDimAT}). The first
level (from below) corresponds to the scaling dimension $x_\sigma=1/8$ of the
magnetization density. Because a magnetization can be defined for each Ising
replica, this level is two-fold degenerate as can be seen from the excellent
superposition of the symbols $+$ and $\times$ (giving the impression of $\ast$
symbols). The second level corresponds to the scaling dimension $x_{\sigma\tau}$
of the polarization density. It is not degenerated and represented with $\ast$
symbols. The third level corresponds to the scaling dimension $x_\varepsilon
=2-y_t$ of the energy density. A systematic deviation from the expected
value Eq.~(\ref{ScalDimAT}) is observed for $J_1\gtrsim 2$, i.e. close to the
4-state Potts point. Finally, the fourth level, corresponding to an exponent
$x_\sigma+1$, can be interpretation as the first descendent of the magnetization
density. It is therefore two-fold degenerate too. The deviation from the
exact value is larger, of the order of 10 to 20\%, and results from a too
small number of states kept at each Tensor Renormalization.
\\

The estimates of scaling dimensions with Open Boundary Conditions are plotted
on the right subplot of Fig~\ref{fig3}. The first level corresponds to
the scaling dimension $x^s_\sigma$ of the surface magnetization density
and is two-fold degenerate. The agreement with the expected value
Eq.~(\ref{ScalSurfDimAT}) is very good, except at the close vicinity of
the 4-state Potts point. The second level is associated to
the surface polarization density. The agreement with the exact value
$x_{\sigma\tau}^s=1$ is excellent for small couplings $J_1$ but a systematic
deviation is observed already for $J_1\gtrsim 2$. The third level is the
first descendent of the surface magnetization density.

\begin{figure}
    \centering
    \includegraphics[width=0.49\textwidth]{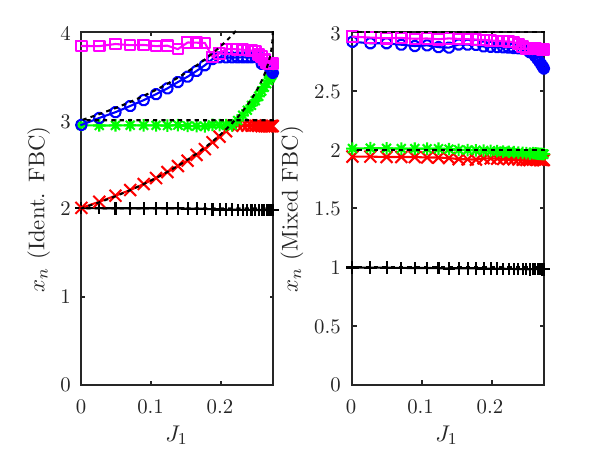}
    \caption{Estimates of surface scaling dimensions of the Ashkin-Teller
    model with Fixed Boundary Conditions (Identical on the left subplot
    and Mixed on the right one) versus the coupling $J_1$. The dashed lines
    are the exact values.}
    \label{fig4}
\end{figure}

With Fixed Boundary Conditions (FBC), only primary operators and their descendants
that are compatible with the BCs are expected to appear in the expansion of
the partition function. Our estimates of their scaling dimensions are plotted
on Fig.~\ref{fig4}. On the left figure, the spins of the two Ising replicas are
imposed to be in the same state on both left and right surfaces of the strip
(Identical FBC). The scaling dimensions are in agreement with $2$ (descendent
of the identity), $4x_\sigma^s$, $3$, $4x_\sigma^s+1$, $4$.
Note the scaling dimension $4x_\sigma^s$ is indeed found in some
irreducible representations of Virasoro algebra~\cite{Gehlen86} and
expected to occur for OBC too. We do not observe it in this case because
it would be above the levels that we computed. On the right subplot of
Fig.~\ref{fig4}, the two Ising replicas are imposed to be
in different states on the right surface of the strip only (Mixed FBC).
Only integer scaling dimensions are observed. They can be associated
to descendants of the identity.

\section{Surface critical behavior of the $J1-J2$ model}
We consider a frustrated two-layer Ising model with Hamiltonian
    \ba
    -\beta H&=&J_2\sum_{i,j}\sigma^A_{i,j}\big[
    \sigma^A_{i+1,j}+\sigma^A_{i,j+1}\big]\nonumber\\
    &&+J_2\sum_{i,j}\sigma^B_{i,j}\big[
    \sigma^B_{i+1,j}+\sigma^B_{i,j+1}\big]\nonumber\\
    &&+J_1\sum_{i,j}\sigma^A_{i,j}\big[
    \sigma^B_{i+1,j}-\sigma^B_{i,j+1}\big]\nonumber\\
    &&+J_1\sum_{i,j}\sigma^B_{i,j}\big[
    \sigma^A_{i+1,j}-\sigma^A_{i,j+1}\big]
    \label{Eq2}
    \ea
introduced in~\cite{Chatelain}.
The intra-layer coupling $J_2$ is ferromagnetic while the inter-layer
one $J_1$ is ferromagnetic in one direction of the lattice and
anti-ferromagnetic in the other one. The phase diagram of the
model is expected to be symmetric under the sign reversal of $J_1$
since it is equivalent to a rotation by $90^\circ$ of the lattice.
It is also symmetric under the sign reversal of both couplings,
i.e. $(J_1,J_2)\rightarrow (-J_1,-J_2)$, since the partition function
is invariant when one concomitantly introduces staggered spins
    \be \sigma_{i,j}^{A,B}\longrightarrow (-1)^{i+j}\sigma_{i,j}^{A,B}.\ee
Combining these two symmetries yields an additional symmetry of the
phase diagram under the sign reversal of $J_2$. Finally, the
transformation
    \be \big(\sigma^A_{i,j},\sigma^B_{i,j}\big)
    \rightarrow (-1)^i\times\left\{\begin{aligned}
    &\big(\sigma^A_{i,j},\sigma^B_{i,j}\big),\quad i+j {\rm\ even},\\
    &\big(\sigma^B_{i,j},\sigma^A_{i,j}\big),\quad i+j {\rm\ odd}
    \end{aligned}\right.        \label{ExchJ1J2}\ee
leaves the partition function invariant if the two couplings are
exchanged. Note that none of these transformations can remove the
frustration of the model.

\subsection{Determination of the critical line}
The phase diagram of the frustrated two-layer Ising model Eq.~(\ref{Eq2})
has been previously determined in~\cite{Chatelain}. A new approach is proposed
in this section. Because of the symmetries discussed above, the study
will be restricted to the region $0<J_1<J_2$ in the following.
\\

In~\cite{Chatelain}, the location of the critical line was estimated
from the behavior of the central charge. The later was extracted from
the Finite-Size Scaling of the free energy density. According to Zamolodchikov
theorem~\cite{Zamo}, the effective central charge is expected to increase
along the renormalization group flow, and therefore to reach a maximum
at the critical point. Accurate critical couplings could be obtained
only for $J_2\lesssim 1/3$. When moving away from the Ising point,
the fit of the free energy density becomes more and more difficult.
In this work, the critical line is determined by studying the Finite-Size
Scaling of the free energy of a domain wall. The latter is obtained as the
difference $\Delta F$ of the free energies of two systems with respectively
identical and mixed FBC. In the paramagnetic phase, the effect of the
BCs is felt by the system only over a distance $\xi$ corresponding to the
correlation length. In the thermodynamic limit, the interface tension
vanishes. In the ordered phase, the mixed BC favors the presence of
a domain wall so that the difference $\Delta F$ of the free energies
of the two systems diverges for an infinite strip.

\begin{figure}
    \centering
    \includegraphics[width=0.49\textwidth]{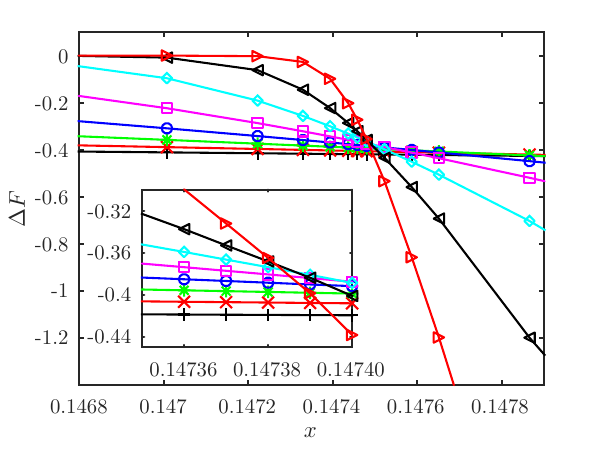}
    \caption{Free energy difference $\Delta F$ of the two-layer Ising model
    with different FBCs at $J_0=0.3042435$ versus the variable
    $x={1\over 2}(J_2-J_1)$. The different curves corresponds to different
    numbers of BTRG iterations, \correction{and therefore different lattice
    sizes. The steepest red curve correponds to the largest lattice size}.
    The data shown in the main plain were computed with $\chi=32$ states
    while in the inset, the data were obtained with $\chi=64$ states.}
    \label{fig8}
\end{figure}

To determine the critical line, a first sweep was performed over the plane
$(J_1,J_2)$ along lines perpendicular to the diagonal $J_1=J_2$, i.e.
we fixed $J_0={1\over 2}(J_1+J_2)$ and varied $x={1\over 2}(J_2-J_1)$.
Preliminary calculations were systematically performed with $\chi=16$ states
in the BTRG algorithm. A first estimate of the critical line was refined
with additional calculations with $\chi=32$ and then $64$ states.
As can be observed on Fig.~\ref{fig8}, the free energy difference
$\Delta F$ displays two distinct behaviors: in the paramagnetic phase,
$\Delta F$ decreases with the number of iterations, and therefore the
size of the system, while in the ordered phase, $\Delta F$ diverges.
The critical point can be estimated as the crossing point of the curves
corresponding to two successive numbers of iterations. In the following,
results are presented for the two largest numbers of iterations.
The critical line was determined up to an accuracy of ${\cal O}(10^{-5})$.
Two sets of BCs were considered. In the first one, the spins were fixed
at the left boundary in the state $(\sigma^A,\sigma^B)=(\uparrow,\uparrow)$
and at the right boundary in the state $(\uparrow,\uparrow)$ (identical BC)
or in $(\downarrow,\downarrow)$ (mixed BC). The second set of FBC was
defined as spins at the left boundary in the state $(\sigma^A,\sigma^B)
=(\uparrow,\downarrow)$ and at the right boundary in the state
$(\uparrow,\downarrow)$ (identical BC) or in $(\downarrow,\uparrow)$
(mixed BC). Because the number of iterations, and therefore the lattice
size, is finite, these two sets of BCs yield slightly different estimates
of the critical line. The difference between them is at most
$2.10^{-5}$.

\begin{figure}
    \centering
    \includegraphics[width=0.49\textwidth]{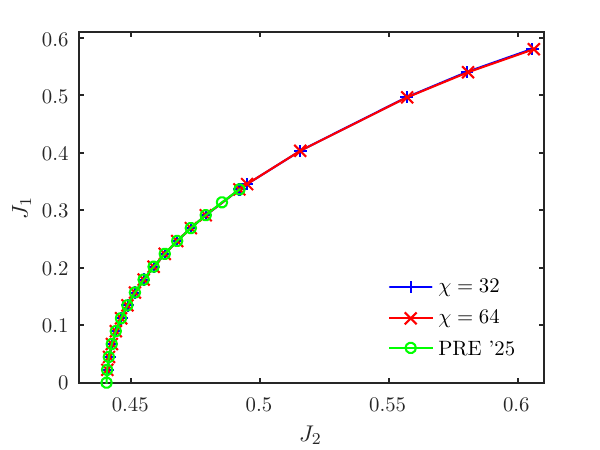}
    \caption{Critical line in the region $0<J_1<J_2$. The blue $+$ symbols
    corresponds to $\chi=32$ states, the red $\times$ symbols to $\chi=64$ states
    and the green $\circ$ symbols were obtained in~\cite{Chatelain}.}
    \label{fig9}
\end{figure}

The critical line is presented on Fig.~\ref{fig9}. For comparison, the
estimate with only $\chi=32$ states in the BTRG algorithm is also plotted.
The difference with the estimate for $\chi=64$ states is of order of
${\cal O}(10^{-4})$. The data estimated in~\cite{Chatelain} as the
maximum of the effective central charge are also plotted. Again, the
difference is of order of ${\cal O}(10^{-4})$.

\subsection{Bulk scaling dimensions}
As for the Ashkin-Teller, the scaling dimensions of the primary operators
and their descendants of the two-layer Ising model were determined by
estimating the largest eigenvalues of the transfer matrix and applying
the gap-exponent relation.
\\

\begin{figure}
    \centering
    \includegraphics[width=0.49\textwidth]{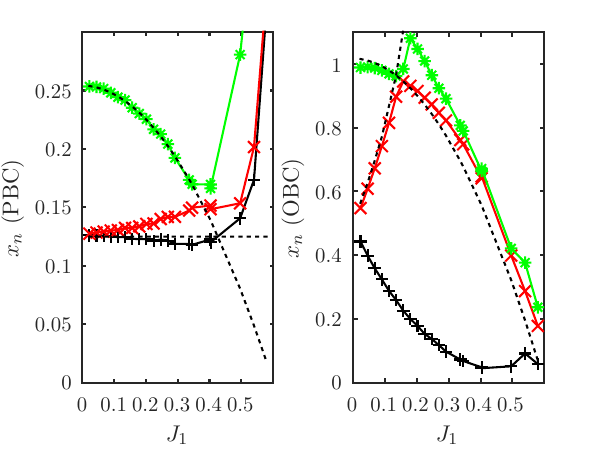}
    \caption{Estimates of the bulk (left) and surface (right) critical
    dimensions of the frustrated two-layer Ising model versus the coupling $J_1$.
    The dashed lines are the conjectured values.}
    \label{fig10}
\end{figure}

The bulk scaling dimensions are plotted on the left plot of Fig.~\ref{fig10}.
They are less accurate than in~\cite{Chatelain} because the calculations
were performed at the same time for a periodic transfer matrix and with
Open or Fixed BCs. Because of the additional computational cost, the number
of states in the BTRG algorithm was limited to $\chi=64$. Nevertheless,
the lowest scaling dimension remains close to the exact value $x_\sigma=1/8$,
as for the Ashkin-Teller model. The third scaling dimension
displays a behavior with $J_1$ that is similar to $x_{\sigma\tau}$
associated to the polarization density of the Ashkin-Teller model.
However, the main discrepancy is that the second scaling dimension was
expected to be degenerate with the first one while a systematic deviation
is observed with $J_1$. By fitting the scaling dimensions with the inverse
of the number of BTRG iterations, as discussed previously for the AT model,
a good agreement of the two first scaling dimensions $1/8$ is recovered.
However, the higher scaling dimensions are noisier so this approach will
not be employed in the following.

A parabolic fit of the numerical estimate of $x_{\sigma\tau}$ with $J_1$
was performed (plotted as a dotted curve on Fig.~\ref{fig10}). From this fit,
the parameter $y$ is estimated from Eq.~(\ref{ScalDimAT}). The expected
dependence of the scaling dimensions $x_\varepsilon$ and $x_\sigma^s$ on $J_1$
can then be deduced from Eq.~(\ref{ScalDimAT}) and Eq.~(\ref{ScalSurfDimAT})
respectively. The parabolic fit of $x_{\sigma\tau}$ also allows to locate the
tricritical point at $J_1\simeq 0.425$, i.e. as the point where $x_{\sigma\tau}
=1/8$ and the model reduces to the 4-state Potts model.

\correction{
For larger values of $J_1\gtrsim 0.425$, the bulk transition is expected
to be of first-order. In this regime, the estimates of the exponents are
seen to increase rapidly with $J_1$ on the right plot of Fig.~\ref{fig10}.
At a first-order phase transition, correlation functions decay
exponentially fast so the gaps $E_n-E_0$ of the transfer matrix remain
finite at the transition. The estimates of the exponents computed
with Eq.~(\ref{GapExponentPBC}) are expected to diverge linearly with
the lattice size $L$ and are therefore meaningless.}

\subsection{Surface scaling dimensions}
With OBC (right plot of Fig.~\ref{fig10}), two new levels, with no equivalent
in the AT model, are present. Both starts at $J_1=0$ from the value $x_\sigma^s
=1/2$ of the scaling dimension of the surface magnetization of the Ising model.
While the 4-spin interaction of the Ashkin-Teller model does not lift the
degeneracy of the two lowest boundary modes, the inter-layer spin-spin coupling
of the F2LIM leads to a level splitting increasing rapidly with
the marginal coupling $J_1$. The degeneracy in the AT model is caused by
the ${\mathbb Z}_2$ symmetry of the Hamiltonian Eq.~(\ref{Eq1})
under the spin reversal of a single Ising replica
    \be (\sigma_{ij}^A,\sigma_{ij}^B)\longrightarrow
    (-\sigma_{ij}^A,\sigma_{ij}^B).         \label{SpinReversal}\ee
The 4-spin inter-layer coupling does not break this symmetry because the energy
is even under the spin reversal of an Ising replica. In contrast, in the F2LIM,
the Ising replicas are coupled by spin-spin interactions, either ferromagnetic
or anti-ferromagnetic. In the bulk, the ${\mathbb Z}_2$ symmetry is present
because of the precise balance of ferromagnetic and anti-ferromagnetic couplings.
The Hamiltonian of the F2LIM is indeed invariant under the spin reversal of an
Ising replica Eq.~(\ref{SpinReversal}) if it is followed by a $90^\circ$
rotation of the lattice. However, the different boundaries of the system are
not equivalent: on two of them, the inter-layer couplings are all ferromagnetic
while on the two others, they are all anti-ferromagnetic. As a consequence, the
symmetry under spin-reversal of a single Ising replica is broken at the boundaries
and the degeneracy of the scaling dimensions is lifted.
\\

As can be seen on Fig.~\ref{fig10} (right), the lowest scaling dimension becomes
very small after the tricritical point. The second one increases with $J_1$
and crosses the next scaling dimension at $J_1\simeq 0.1$ and the following
one at $J_1\simeq 0.2$. Besides this point, it goes out of the range
of eigenvalues that we measured. We observed that these two lowest (up to $J_1
\simeq 0.1$) scaling dimensions $x_1^s$ and $x_2^s$ are related by
	\be x_2^s\simeq {1\over 4x_1^s}.			\label{EqSym}\ee
On Fig.~\ref{fig10}, one can see the good agreement of the second level
with the dotted line computed from the numerical data of $x_1^s$ as
$1/4x_1^s$. Eq.~(\ref{EqSym}) is similar to the duality relation between
electric scaling dimensions $x_e(n)=n^2/2g$ and magnetic ones $x_m(m)=m^2g/2$
in the Coulomb gas with $n=m=1$. However, the $g$ coupling that can be
computed from the numerical estimate of $x_1^s$ is significantly smaller
than the expected value ${2\over x_\varepsilon}=x^s_\sigma$ for the AT
model and therefore the F2LIM~\cite{Saleur}. The third (up to $J_1\simeq 0.1$)
scaling dimension is in good agreement with the value $x_\varepsilon
=4x_{\sigma\tau}$ of the AT model in the bulk.

\begin{figure}
    \centering
    \includegraphics[width=0.49\textwidth]{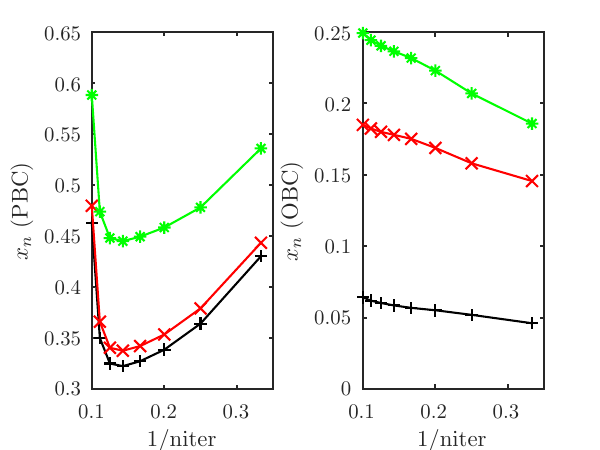}
    \caption{\correction{
    Estimates of the bulk (left) and surface (right) scaling dimensions
    of the frustrated two-layer Ising model versus the inverse
    of the number of iterations of the BTRG algorithm at the point
    $J_1=0.5802040$ (in the regime of first-order bulk transition).}}
    \label{fig15}
\end{figure}

\correction{
On the right plot of Fig.~\ref{fig15}, one can observe for a point in
the regime of first-order bulk phase transition that the numerical
estimates of the surface scaling dimensions evolve smoothly (close to linearly)
with the inverse of the number of BTRG iterations. In constrast, the
bulk scaling dimensions diverge at large number of iterations. While
the latter could result from the finite number of states $\chi$
kept in the Renormalization process and not necessarily from a first-order
bulk transition, the former is a clear evidence of a continuous surface
phase transition. Indeed, as already mentioned, the gap would remain finite
at a first-order phase transition so the application of the
gap-exponent relation Eq.~(\ref{GapExponentPBC}) would lead to estimates
of the scaling dimensions diverging at large number of iterations.
However, the possibility of a first-order surface transition
with a correlation length at the transition larger that the largest lattice
size cannot be excluded.}
\\

\correction{As shown on the right plot of Fig.~\ref{fig10}, the first
surface scaling dimension approaches zero when $J_1$ grows. This value
is remarkably different from the surface magnetic exponent of the
$q$-state Potts model that takes values close to 3 in the first-order
regime~\cite{Igloi}. Surprinsingly, the second and third scaling dimensions
seems to become degenerate in the regime of first-order bulk transition
and are in agreement with the parabolic fit of $x_\varepsilon=4x_{\sigma\tau}$
of the AT model.}

\subsection{Fixed Boundary Conditions}
\begin{figure}
    \centering
    \includegraphics[width=0.49\textwidth]{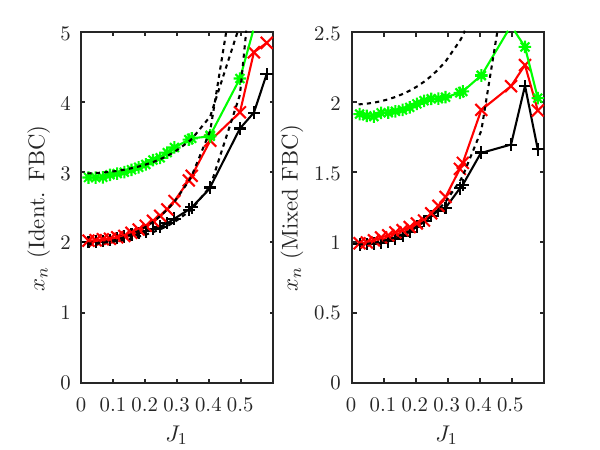}
    \caption{Estimates of surface scaling dimensions of the two-layer Ising
    model with the first set of Fixed Boundary Conditions (Identical on the
    left and Mixed on the right) versus the coupling $J_1$. The dashed
    lines are the conjectured values.}
    \label{fig11}
\end{figure}

With the first set of BCs and identical FBC, the two lowest scaling dimensions
are in good agreement with $2x_\sigma^s+1$ and to $4x_\sigma^s$
(Fig.~\ref{fig11}). They are degenerated at the Ising point $J_1=0$. The third
one is in good agreement in the second-order regime with the scaling dimension
$2x_\sigma^s+2$ of the next descendent. With mixed scaling dimensions, the two
lowest scaling dimensions are in good agreement with $2x_\sigma^s$ in the
second-order regime. Again, the third scaling dimension corresponds probably to
$2x_\sigma^s+1$, i.e. its first descendent. In contrast to the AT model,
integer scaling dimensions, i.e. descendent of the identity, are not present
in the spectrum.

\begin{figure}
    \centering
    \includegraphics[width=0.49\textwidth]{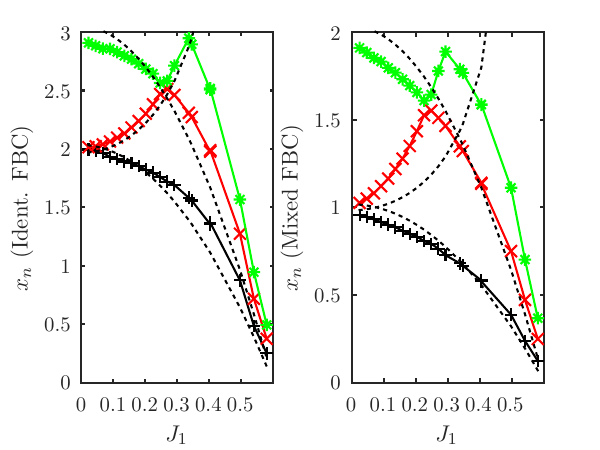}
    \caption{Estimates of surface scaling dimensions of the two-layer Ising
    model with the second set of Fixed Boundary Conditions (Identical on the
    left and Mixed on the right) versus the coupling $J_1$. The dashed lines
    are the conjectured values.}
    \label{fig12}
\end{figure}

With the second set of BCs and identical FBC, the lowest scaling dimension
is in good agreement with the exponent $2x_\varepsilon$ ($=8x_{\sigma\tau}$)
of the AT model (Fig.~\ref{fig12}). The second scaling dimension, which becomes
the third one for $J_1\gtrsim 0.25$, is close to $4x_\sigma^s$. The third scaling
dimension (up to $J_1\simeq 0.25$) is in good agreement with $3x_\varepsilon$.
It would therefore corresponds to a new primary operator. A more probable
scenario is that this scaling dimension is $2x_\varepsilon+1$ and is therefore
the first descendent of the lowest state. With mixed BC, the spectrum quite
similar: the lowest scaling dimension is compatible with $x_\varepsilon$ and
the third one with $2x_\varepsilon$. More probably, the latter corresponds to
$x_\varepsilon+1$. However, we have no interpretation for the second
scaling dimension. As can be seen on Fig.~\ref{fig12}, it is compatible
with $2x_\sigma^s=1$ at the Ising point but then increases too fast with $J_1$
($2x_\sigma^s$ is plotted as a dotted line).

\begin{table*}
\caption{\label{Tab1}Scaling dimensions of the Ashkin-Teller model.}
\begin{ruledtabular}
\begin{tabular}{l|llll}
Level & PBC & OBC & Id. FBC & Mix. FBC \\
\hline
1 & $x_\sigma={1\over 8}$ ($\times 2$) & $x_\sigma^s=1/2x_\varepsilon$
($\times 2$) & 2 & 1 \\
2 & $x_{\sigma\tau}$ & 1 & $4x_\sigma^s$ & 2 ($\times 2$)\\
3 & $x_\varepsilon=4x_{\sigma\tau}$ & $x_\sigma^s+1$ ($\times 2$) & 3 & 3 ($\times 2$)\\
4 & 1 & - & $4x_\sigma^s+1$ & - \\
\end{tabular}
\end{ruledtabular}
\end{table*}

\begin{table*}
\caption{\label{Tab2}Scaling dimensions of the frustrated two-layer Ising model.}
\begin{ruledtabular}
\begin{tabular}{l|llllll}
Level & PBC & OBC & Id. FBC & Mix. FBC & Id. FBC & Mix. FBC \\
\hline
1 & $x_\sigma={1\over 8}$ ($\times 2$) & $x_1^s$ &
$2x_\sigma^s+1$ & $2x_\sigma^s$ ($\times 2$) &
$2x_\varepsilon$ & $x_\varepsilon$   \\
2 & $x_{\sigma\tau}$ & $x_2^s=1/4x_1^s$ &
$4x_\sigma^s$ & $2x_\sigma^s+1$ &
$4x_\sigma^s$ & $??$ \\
3 & - & $x_\varepsilon$ & $2x_\sigma^s+2$ & - &
$3x_\varepsilon$\ {\rm or}\ $2x_\varepsilon+1$ &
$2x_\varepsilon$\ {\rm or}\ $x_\varepsilon+1$ \\
\end{tabular}
\end{ruledtabular}
\end{table*}

\section*{Conclusions}
\correction{
We have presented a new extension of the BTRG algorithm to systems
with boundaries, in the spirit of Refs.~\cite{Ilino,Ilino2} for the HoTRG
algorithm. While previous applications were limited to Ising and Potts models,
our implementation was applied to both the Ashkin–Teller (AT) model and
the Frustrated two-layer Ising model (F2LIM) introduced in~\cite{Chatelain},
which belong to the same universality class in the bulk. From a numerical
perspective, the BTRG approach provides stable estimates of both bulk
and surface scaling dimensions while previous attempts to study the critical
behavior of the $J_1-J_2$ model, sharing the same scaling limit as the F2LIM,
with Tensor-Network techniques failed~\cite{Li1,Yoshiyama,Gangat}.
We have shown that the optimal parameter of the BTRG model remains
$k=-1/2$, even in presence of frustration. The method directly yields
the lowest conformal scaling dimensions from the transfer-matrix gaps,
offering a viable alternative to Monte Carlo simulations.
}
\\

\correction{
From the physical point of view, the BTRG calculations yield an estimate
of the magnetic surface exponent $x_\sigma^s$ of the AT model along
its critical line in agreement with previous estimates~\cite{Gehlen86,
Gehlen87,Alcaraz87}, confirming the accuracy of the method. The conformal
spectrum with FBCs is determined. Only descendants of the identity are
compatible with identical boundaries whereas a new operator of scaling
dimension $4x_\sigma^s$ appears with mixed boundaries.
}
\\

\correction{
The conjectured scaling dimensions for the AT and F2LIM are summarized
in Tables~\ref{Tab1} and~\ref{Tab2}. Despite sharing the same bulk universality
class, the boundary conformal spectrum of these two models is different
suggesting that the F2LIM realizes a distinct boundary conformal field theory
representation of the same bulk $c=1$ theory.
In particular, the two-fold degeneracy of the lowest surface scaling
dimension found in the AT model is lifted due to the anisotropy of the F2LIM.
Neverthess, most of the scaling dimensions of the spectrum with both Open
and Fixed BCs can be associated either to primary operators of the AT model
or their descendants with the appropriate dependence of the parameter $y$
with the interlayer coupling $J_1$ of the F2LIM.
}

\section*{Acknowledgments}
This work was supported by the french ANR-PRME UNIOPEN grant (ANR-22-CE30-0004-01).

\end{document}